\pdfoutput=1
\documentclass[conference]{IEEEtran}
\IEEEoverridecommandlockouts

\usepackage[utf8]{inputenc}
\usepackage[T1]{fontenc}

\usepackage{cite}
\usepackage{amsmath,amssymb,amsfonts}
\usepackage{textcomp}
\usepackage{graphicx}
\usepackage{xcolor}
\usepackage{siunitx}
\usepackage{booktabs}
\usepackage[inline]{enumitem}
\usepackage{tikz}
\usetikzlibrary{shapes,arrows,positioning}

\usepackage{listings}

\usepackage{xurl}
\usepackage[hidelinks]{hyperref}
\title{Iola Walker: A Mobile Footfall Detection System for Music Composition}
\author{\IEEEauthorblockN{William B. James}
\IEEEauthorblockA{Audi8 Labs\\
New York, NY USA}}

\begin{document}
\maketitle

% -------------------- Abstract ---------------------------------
\begin{abstract}
This outing is part of a larger music technology research project. The objective is to find a way to enhance music using hardware and software. This is the documentation for the Whimsical first part of the research project: it's an android app that detects a wearer's footfalls by running live inference on an LSTM. The system works by getting data from an Mbient Labs IMU to a mobile app over bluetooth. After you move the .csv file to a computer with a GPU, you can use the python code to train an LSTM on that data. You then export the LSTM to the android app and can begin detecting footfalls. Feel free to download and experiment with the code. It's meant to be read and improved upon by you and your LLM codewriter of choice! \url{https://github.com/willbjames/iolawalker}
\end{abstract}

% ===============================================================

\section{Introduction: The Starting Point For A New Music Composition and Listening System}
\par The project is dubbed "iola walker" in reference to a common polyrhythm, the hemiola. In the imagined music composition system, musicians record songs with a walking listener in mind, imagining a range of walking paces while playing rhythmic parts. The same passages in a musical composition might be recorded multiple times with a variety of underlying pulses. In the associated interactive listening experience, a listener dons a foot-mounted inertial measurement unit and goes for a walk. The Iola Walker mobile app estimates footfall timing from the IMU over bluetooth.
\par Novel music composition gestures might be derived and implemented using this system.
Perhaps a playback system could select the version of a recorded musical passage with underlying polyrhythms or "pulse" closest to the listener's walking pace. Perhaps the crash at the end of a drum fill could be timed to line up precisely with a footstep. Perhaps the next passage of a song could be determined according to a detected change in walking speed that may have been induced by the previous section.
\par
Information about predicted future footfalls might be passed from the footfall detection process to the audio playback code running in a render process. This design allows for relevant signals, such as footfalls, to be incorporated into music with zero latency.
\par This paper documents the process of training a model to detect a walking listener's footfalls from real-time signal data, as well as exporting that model to work with low latency on an android application. The model was trained on signals from an Mbient Labs foot-mounted IMU \cite{mbientlabs} collecting data at 200 Hertz, with the approximate for footfalls annotated by pressing the volume up button on a BLU JL8 android device \cite{bluJ8L} when the listener's foot hits the ground. To collect training data, I walked around my neighborhood while clicking the volume up button on an android mobile phone each time my foot hit the ground. I tried several models, and found that an LSTM attained the best performance.

\section{Related Works}
\par 
The signal from a person's gait has been identified by other teams as a strong candidate for music technology research. But prior research on gait-modulated music ended up focusing on what the wearer was doing more than the music being played.
\par
The Djogger research project was the first to explore this technology in order to help enhance exercise in 2006 \cite{moens2014djogger}. Then several startups in the 2010s including Weav and Rock My Run, created a system for running workouts by changing the speed of songs played to match running pace \cite{weav2017forbes,rockmyrun2014study}. Most recently, in the 2020s, Medrhythms, in a partnership with Universal Music Group, has obtained FDA approval for its InTandem music therapy technology to be prescribed for people recovering from a stroke. \cite{medrhythmsFDA2023,medrhythmsRx2023}.
\par
The method to detect the footfall signal so that it is usable in real time is similar to prior LTSM methods documented as being highly accurate in non-music literature \cite{vidyarani2025,jeon2024,chen2025} but where no open source code for real time inference has been published.

\section{Mobile Data Acquisition Pipeline}
The shoe-mounted MbientLab IMU streams 200~Hz tri-axial acceleration to
the phone via Bluetooth LE. An Android service wraps the sensor API and
forwards raw samples to \texttt{SensorDataManager}, whose responsibilities
are:

\begin{itemize}
  \item \textbf{Bounded queue} (5~s capacity, \SI[parse-numbers=false]{>99}{\percent} non-blocking writes).
  \item \textbf{Background thread} that drains the queue, batches samples,
        and flushes to a \SI{64}{\kilo\byte} buffered CSV writer every
        \SI{1}{\second} or 1000 samples.
  \item \textbf{Footfall annotation}---each volume-up press inserts a
        high-priority event with the tag \texttt{FOOTFALL}.
  \item \textbf{Session management}---files are timestamped and recycled
        automatically to prevent data loss during long recordings.
\end{itemize}

The resulting CSVs constitute the \texttt{train200hz.csv} and
\texttt{test200hz.csv} datasets used in the Python pipeline below.

\section{Conv-BI-LSTM Model Definition}
In this section we break down each component of the \texttt{ConvLSTMGlobal} architecture in detail.

\subsection{1D Convolutional Layers}
The first two layers are 1D convolutions:
\begin{itemize}
  \item \texttt{nn.Conv1d(in\_ch, 16, kernel\_size=3, padding=1)}
  \item \texttt{nn.ReLU()}
  \item \texttt{nn.Conv1d(16, in\_ch, kernel\_size=3, padding=1)}
  \item \texttt{nn.ReLU()}
\end{itemize}

\subsection{Bidirectional LSTM}
Next, we feed the convolved features into a bidirectional Long Short-Term Memory model (Bi-LSTM):

\begin{itemize}
  \item \textbf{LSTM cells} maintain an internal memory that can capture how sensor readings evolve over time.
  \item \textbf{Bidirectional}: Each window is processed twice---once from start to end and once in reverse.
  \item \(\texttt{hidden\_size}=hid\) controls the capacity of this memory.
\end{itemize}

The output of the Bi-LSTM at each time step is a vector of length \(2\times hid\).

\subsection{Two Heads for Prediction}
After the LSTM, we attach two small heads that turn the LSTM outputs into actual footfall predictions:

\begin{itemize}
  \item \textbf{Sequence head} (\texttt{seq\_head})
  \item \textbf{Global head} (\texttt{global\_head})
\end{itemize}

\section{Live Mobile Inference Pipeline}
The Android application embeds the trained Conv--LSTM model as a TorchScript asset and runs all inference on a dedicated high-priority thread.

\subsection{Streaming Feature Buffer}
As accelerometer data arrives:
\begin{enumerate}
  \item Wraps each sample in a small \texttt{dataWindow} deque entry.
  \item Maintains at most \texttt{windowSize} entries.
  \item Posts \texttt{runModelInference()} once \texttt{inferenceStride} samples accumulate.
\end{enumerate}

\subsection{Model Inference}
Inside \texttt{runModelInference()}:
\begin{itemize}
  \item Flatten \& normalize.
  \item TorchScript forward pass.
  \item Sigmoid conversion.
\end{itemize}

\subsection{Decision Logic and Smoothing}
Footfall events require:
\begin{itemize}
  \item \(p_t > \texttt{sampleThreshold}\)
  \item \(P_{\mathrm{win}} > \texttt{windowThreshold}\)
  \item Majority filter
  \item Refractory period
\end{itemize}

\subsection{Runtime Parameter Tuning via Sliders}
This is a noisy little guy as of the time of this writing! Sliders adjust thresholds, stride, majority filter, and minimum interval.

\section{Current System Behavior}

As of the date of this release, the mobile app outputs a ding when a footfall is detected. As of now the system struggles a lot with precision. It picks up a lot of false positives! With that said, it's recall appears to be surprisingly decent. It's a work in progress. The reader is invited to download the code and experiment with it! The code is in a single folder with no large files or datasets for ease of compartmentalized LLM codewriter iteration.

\section{Conclusions and Future Directions}
\par One thing that became clear in doing this research is that music strongly skews towards being a one-way, unilateral form of communication. Specifically, it's often going to be the most compelling move from an artistic standpoint to capture exactly what the person played in the moment. There are not many existing examples of highly effective, interactive musical compositions. This makes it a worthwhile challenge!

\end{document}